\documentclass{article}%
\usepackage{graphicx}
\usepackage{amsmath}
\usepackage{amsfonts}
\usepackage{amssymb}%
\begin{document}

\author{Antony Valentini\\Augustus College}

\begin{center}
{\LARGE Dynamical Origin of Quantum Probabilities}

\bigskip

\bigskip\bigskip

\bigskip

\bigskip

Antony Valentini\footnote{email: avalentini@perimeterinstitute.ca}

\textit{Perimeter Institute for Theoretical Physics,}

\textit{35 King Street North, Waterloo, Ontario N2J 2W9,
Canada.\footnote{Corresponding address.}}

\textit{Augustus College,}

\textit{14 Augustus Road, London SW19 6LN, England.\footnote{Permanent
address.}}

\bigskip

Hans Westman\footnote{email: hawe@fy.chalmers.se}

\textit{Department of Astronomy and Astrophysics,}

\textit{Chalmers University of Technology,}

\textit{41296 G\"{o}teborg, Sweden}.

\bigskip
\end{center}

\bigskip

\bigskip

\bigskip

We study the origin of the Born probability rule $\rho=|\psi|^{2}$ in the de
Broglie-Bohm pilot-wave formulation of quantum theory. It is argued that
quantum probabilities arise dynamically, and have a status similar to thermal
probabilities in ordinary statistical mechanics. This is illustrated by
numerical simulations for a two-dimensional system. We show that a simple
initial ensemble with a non-standard distribution $\rho\neq|\psi|^{2}$ of
particle positions evolves towards the quantum distribution to high accuracy.
The relaxation process $\rho\rightarrow|\psi|^{2}$ is quantified in terms of a
coarse-grained \textit{H}-function (equal to minus the relative entropy of
$\rho$ with respect to $|\psi|^{2}$), which is found to decrease approximately
exponentially over time, with a time constant that accords with a simple
theoretical estimate.

\bigskip\bigskip

\bigskip

\bigskip

\bigskip

\bigskip

\bigskip

\bigskip

\bigskip

\bigskip

\bigskip

\bigskip

\bigskip

\bigskip

\bigskip

\bigskip

\bigskip\bigskip

\bigskip

\section{Introduction}

In the de Broglie-Bohm pilot-wave formulation of quantum theory [1--10], the
quantum state of an individual system is supplemented by a deterministic
trajectory in configuration space. It is usually assumed that an ensemble of
systems with wave function $\psi(q,t)$ (in configuration space) has
configurations $q$ distributed according to the Born probability rule
$\rho(q,t)=|\psi(q,t)|^{2}$. More precisely, it is assumed that $\rho
(q,0)=|\psi(q,0)|^{2}$ at some initial time $t=0$, the equations of motion
then guaranteeing that this quantum distribution is preserved at all later times.

The status of this assumption, that $\rho=|\psi|^{2}$ at some initial time,
has been the subject of debate. Soon after Bohm's first papers on the theory
appeared \cite{Bohm}, Pauli objected that taking a particular distribution
$\rho=|\psi|^{2}$ as an initial condition was unjustified in a fundamentally
deterministic theory \cite{Pauli53}. Similarly, it was pointed out by Keller
\cite{Keller} that $\rho=|\psi|^{2}$ should be derived from the dynamics, in
order for quantum theory truly to emerge as the statistical mechanics of an
underlying deterministic theory.

Despite these criticisms, Bohm had in fact already suggested in his original
papers that the distribution $\rho=|\psi|^{2}$ had a status similar to that of
thermal equilibrium in ordinary statistical mechanics, and he had conjectured
that this distribution could be derived by appropriate statistical-mechanical
arguments. In a subsequent paper, Bohm studied the specific case of an
ensemble of two-level molecules subjected to random external collisions, and
argued that an initial nonequilibrium distribution $\rho\neq|\psi|^{2}$ would
relax to the equilibrium state $\rho=|\psi|^{2}$ \cite{Bohm53}. A general
argument for relaxation was not given, however. A year later, in response to
the criticisms of Pauli and Keller, and motivated by the difficulties in
formulating a relaxation argument for an arbitrary system, Bohm and Vigier
modified the theory by introducing stochastic `fluid fluctuations' that drive
the required process $\rho\rightarrow|\psi|^{2}$ for a general system
\cite{BV54}.

Most subsequent presentations of the original (deterministic) theory have
taken $\rho=|\psi|^{2}$ as an axiom. Usually, it is simply assumed that any
ensemble prepared at time $t=0$ with wave function $\psi(q,0)$ will have a
distribution $\rho(q,0)=|\psi(q,0)|^{2}$ of actual configurations $q$.
Alternatively, $|\psi|^{2}$ has been regarded as the natural measure of
probability or `typicality' for initial configurations of the whole universe
(taking $\psi$ to be the universal wave function), yielding the Born rule for
all subsystems at all times \cite{DGZ92}.

However, a quite general argument for the relaxation $\rho\rightarrow
|\psi|^{2}$ may be framed in terms of an analogue of the classical
coarse-graining \textit{H}-theorem, based on the \textit{H}-function%
\begin{equation}
H=\int dq\ \rho\ln(\rho/|\psi|^{2}) \label{Hfn}%
\end{equation}
(equal to minus the relative entropy of $\rho$ with respect to $|\psi|^{2}$)
\cite{PLA1,AVth,ValIsch}. Like its classical counterpart, this theorem
provides a general mechanism in terms of which one can understand how
equilibrium is approached, while not proving that equilibrium is actually
reached (see Sect. 5 below).

A relaxation timescale $\tau$ -- analogous to the scattering time of classical
kinetic theory -- may be defined in terms of time derivatives of the
coarse-grained $H$-function. For a particle of mass $m$, whose wave function
has a quantum energy spread $\Delta E$, the timescale is estimated to be
\cite{ValIsch}%
\begin{equation}
\tau\sim\frac{1}{\varepsilon}\frac{\hbar^{2}}{m^{1/2}(\Delta E)^{3/2}}
\label{tau}%
\end{equation}
where $\varepsilon$ is the coarse-graining length. One expects $\tau$ to be
the timescale over which there will be a significant approach to equilibrium.
This rough estimate agrees quite closely with numerical simulations for
particles moving in one spatial dimension \cite{ValIsch}.

In this paper, we study the relaxation $\rho\rightarrow|\psi|^{2}$
numerically, for the more realistic case of particles moving in two spatial
dimensions. We shall see that if $\psi$ is a superposition of energy
eigenstates, the de Broglie-Bohm velocity field varies extremely rapidly in
general, particularly in regions where $|\psi|$ is small. The resulting
particle trajectories have a very complicated structure. Starting from a
simple nonequilibrium distribution $\rho_{0}\neq|\psi_{0}|^{2}$ at $t=0$, the
relaxation $\rho\rightarrow|\psi|^{2}$ takes place rather efficiently, on a
timescale of order (\ref{tau}). And a plot of the coarse-grained
\textit{H}-function against time shows an approximately exponential decay,
with time constant of order (\ref{tau}).

We interpret these results as further evidence that, in the pilot-wave
formulation of quantum theory, the Born distribution $\rho=|\psi|^{2}$ should
not be regarded as an axiom. Rather, it should indeed be seen as dynamically
generated, in the same sense that one usually regards thermal equilibrium as
arising from a process of relaxation based on some underlying dynamics.

In Sect. 2, we introduce pilot-wave dynamics for a single particle in a
two-dimensional box. In Sect. 3, we illustrate the character of the
trajectories. In Sect. 4, we introduce a nonequilibrium ensemble of systems,
and show in detail how the ensemble evolves towards $\rho=|\psi|^{2}$. In
Sect. 5 we show how the relaxation may be quantified in terms of the
coarse-grained \textit{H}-function, which is found to decay approximately
exponentially with time, on a timescale of order (\ref{tau}). In Sect. 6, we
discuss the significance of these results, addressing some relevant issues in
the foundations of statistical mechanics.

\section{Pilot-Wave Dynamics in Two Dimensions}

Consider a single particle of unit mass, moving in a potential $V$ in two
spatial dimensions. The system has a configuration $q=(x,y)$, and wave
function $\psi=\psi(x,y,t)$ (assuming a pure state) satisfying the
Schr\"{o}dinger equation%
\begin{equation}
i\frac{\partial\psi}{\partial t}=-\frac{1}{2}\frac{\partial^{2}\psi}{\partial
x^{2}}-\frac{1}{2}\frac{\partial^{2}\psi}{\partial y^{2}}+V\psi\label{Sch}%
\end{equation}
(units $\hbar=1$).

In de Broglie-Bohm theory, the particle traces out a definite trajectory
$(x(t),y(t))$, which is determined by the wave function $\psi$ according to
the de Broglie guidance law%
\begin{equation}
\frac{d\mathbf{x}}{dt}=\operatorname{Im}\frac{\mathbf{\nabla}\psi}{\psi
}=\mathbf{\nabla}S \label{deB}%
\end{equation}
(where $\psi=|\psi|e^{iS}$). Mathematically, $\operatorname{Im}(\mathbf{\nabla
}\psi/\psi)$ is just the ratio of the quantum probability current to the
quantum probability density. Physically, however, $\psi$ is here interpreted
as an objective field (in configuration space), guiding the motion of a
\textit{single} system.

Indeed, so far we have said nothing about probabilities or ensembles, and the
equations (\ref{Sch}) and (\ref{deB}) define a deterministic dynamics for
individual particles. Given the initial wave function $\psi(x,y,0)$,
(\ref{Sch}) determines the wave function $\psi(x,y,t)$ at all times; and given
the initial particle position $(x(0),y(0))$, (\ref{deB}) then determines the
particle trajectory $(x(t),y(t))$ at all times.

For an ensemble of independent particles, each guided by the same wave
function $\psi(x,y,t)$, we may define a density $\rho(x,y,t)$ of actual
configurations $(x,y)$ at time $t$. The guidance equation (\ref{deB}) defines
a velocity field $(\dot{x},\dot{y})$, and for an ensemble of particles with
the same wave function $\psi$ the field $(\dot{x},\dot{y})$ determines the
evolution of any distribution $\rho(x,y,t)$ via the continuity equation%
\begin{equation}
\frac{\partial\rho}{\partial t}+\frac{\partial(\rho\dot{x})}{\partial x}%
+\frac{\partial(\rho\dot{y})}{\partial y}=0 \label{Cont}%
\end{equation}

Because (\ref{Sch}) implies the continuity equation%
\begin{equation}
\frac{\partial\left\vert \psi\right\vert ^{2}}{\partial t}+\frac
{\partial(\left\vert \psi\right\vert ^{2}\dot{x})}{\partial x}+\frac
{\partial(\left\vert \psi\right\vert ^{2}\dot{y})}{\partial y}=0
\label{Contpsi2}%
\end{equation}
for $\left\vert \psi\right\vert ^{2}$, the particular initial distribution
$\rho(x,y,0)=|\psi(x,y,0)|^{2}$ evolves into $\rho(x,y,t)=\left\vert
\psi(x,y,t)\right\vert ^{2}$. The state of `quantum equilibrium' is preserved
by the dynamics.

Note that (\ref{Cont}) determines the evolution of \textit{any} initial
distribution $\rho(x,y,0)$, even if $\rho(x,y,0)\neq|\psi(x,y,0)|^{2}$. Given
$\psi(x,y,0)$, (\ref{Sch}) determines $\psi(x,y,t)$ at all times, and
(\ref{deB}) then determines the velocity field $(\dot{x},\dot{y})$ at all
times. Once the velocity field $(\dot{x},\dot{y})$ is known everywhere,
(\ref{Cont}) may be integrated to yield $\rho(x,y,t)$ at all times, for any
initial $\rho(x,y,0)$.

Before we proceed, it is instructive to compare the above with the analogous
classical Hamiltonian evolution on phase space. Classically, the trajectory
$\left(  q(t),p(t)\right)  $ of an individual system is determined by
Hamilton's equations%
\begin{equation}
\dot{q}=\frac{\partial H}{\partial p},\;\;\;\;\dot{p}=-\frac{\partial
H}{\partial q} \label{Ham}%
\end{equation}
given the initial values $\left(  q_{0},p_{0}\right)  $. These equations
define a velocity field $\left(  \dot{q},\dot{p}\right)  $ on phase space, and
for an ensemble of systems with the same Hamiltonian $H$ the field $\left(
\dot{q},\dot{p}\right)  $ determines the evolution of any distribution
$\rho(q,p,t)$ via the continuity equation%
\begin{equation}
\frac{\partial\rho}{\partial t}+\frac{\partial(\rho\dot{q})}{\partial q}%
+\frac{\partial(\rho\dot{p})}{\partial p}=0 \label{Contcl}%
\end{equation}
which may be rewritten as%
\begin{equation}
\frac{\partial\rho}{\partial t}+\left\{  \rho,H\right\}  =0 \label{Liou}%
\end{equation}
Because $\partial\dot{q}/\partial q=-\partial\dot{p}/\partial p$, a uniform
initial distribution $\rho(q,p,0)=\mathrm{const}.$ (on the energy surface)
remains uniform, $\rho(q,p,t)=\mathrm{const}.$. The state of thermal
equilibrium is preserved by the dynamics. While for a general (non-uniform)
initial state $\rho(q,p,0)$, the evolution $\rho(q,p,t)$ is obtained (in
principle) by integrating (\ref{Liou}).

It is known that in appropriate circumstances, the classical evolution on
phase space defined by (\ref{Liou}) leads to thermal relaxation on a
coarse-grained level (see Sect. 5 below). Less is known about the
corresponding evolution (on configuration space) defined by (\ref{Cont}).

Classically, the canonical example of thermal relaxation (for an isolated
system) is that of a simple initial distribution $\rho(q,p,0)$ (with no
fine-grained microstructure) concentrated on a small region of the energy
surface. Under the Hamiltonian evolution (9), the distribution $\rho(q,p,t)$
will generally develop a complex filamentary structure that spreads out over
the energy surface, so that the distribution approaches uniformity on a
coarse-grained level.

Here, we shall consider an analogous example, of particles in a
two-dimensional box with a simple initial distribution $\rho(x,y,0)\neq
|\psi(x,y,0)|^{2}$ (with no fine-grained microstructure). We shall see, by
numerical integration of (\ref{Cont}), that the ensemble evolves towards the
equilibrium distribution $|\psi|^{2}$ (on a coarse-grained level).

Consider, then, a particle confined to a square box of side $\pi$, with
infinite barriers at $x,\ y=0,\ \pi$. The energy eigenfunctions are%
\begin{equation}
\phi_{mn}(x,y)=\frac{2}{\pi}\sin\left(  mx\right)  \sin\left(  ny\right)
\label{Efns}%
\end{equation}
with energy eigenvalues $E_{mn}=\frac{1}{2}(m^{2}+n^{2})$, where
$m,n=1,2,3,.....$ are positive integers.

As a specific example, we take the initial wave function $\psi(x,y,0)$ at
$t=0$ to be a superposition of the first 16 modes, $m,n=1,2,3,4$, with
amplitudes of equal modulus but randomly-chosen phases $\theta_{mn}%
$:\footnote{For the record, the values of $\theta_{mn}$ used were (to four
decimals) $\theta_{11}=5.1306$, $\theta_{12}=2.0056$, $\theta_{13}=4.1172$,
$\theta_{14}=3.3871$, $\theta_{21}=6.2013$, $\theta_{22}=4.6598$, $\theta
_{23}=1.8770$, $\theta_{24}=4.3033$, $\theta_{31}=4.0145$, $\theta
_{32}=6.1142$, $\theta_{33}=5.4401$, $\theta_{34}=1.9292$, $\theta
_{41}=3.4015$, $\theta_{42}=6.2109$, $\theta_{43}=6.0370$, $\theta
_{44}=5.9159$.}%
\begin{equation}
\psi(x,y,0)=%
%TCIMACRO{\dsum \limits_{m,n=1}^{4}}%
%BeginExpansion
{\displaystyle\sum\limits_{m,n=1}^{4}}
%EndExpansion
\frac{1}{4}\phi_{mn}(x,y)\exp(i\theta_{mn}) \label{psi0}%
\end{equation}
The squared-amplitude $|\psi(x,y,0)|^{2}$ is shown in Fig. 1.

\begin{figure}[h!]
  \begin{center}
    \leavevmode
\resizebox{12cm}{10cm}{
        
    \includegraphics{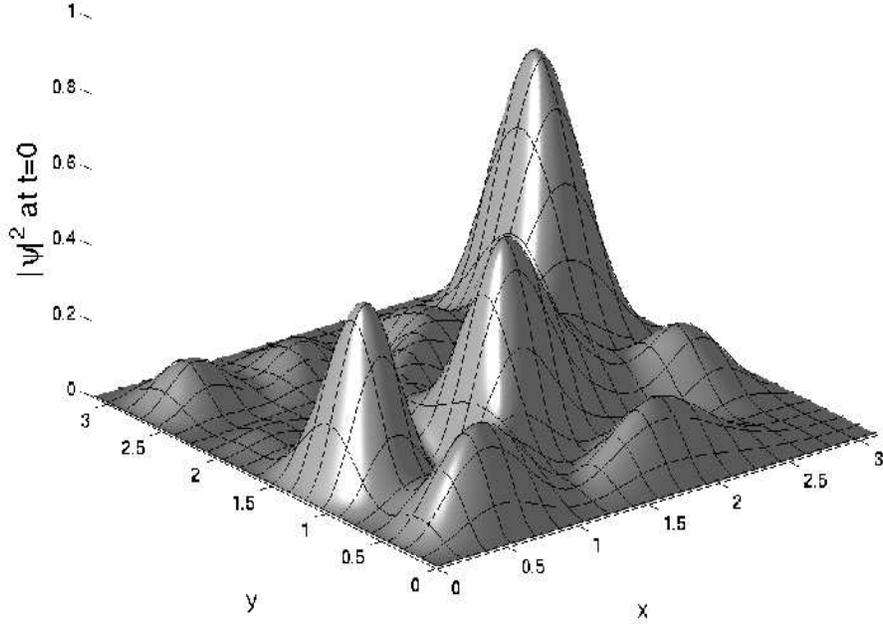}}
    \caption{The squared-amplitude $|\psi(x,y,0)|^{2}$ at $t=0$, for the
specified superposition of the first 16 modes of a two-dimensional box.}
  \end{center}
\end{figure}

From (\ref{Sch}), the wave function $\psi(x,y,t)$ at later times is just%
\begin{equation}
\psi(x,y,t)=%
%TCIMACRO{\dsum \limits_{m,n=1}^{4}}%
%BeginExpansion
{\displaystyle\sum\limits_{m,n=1}^{4}}
%EndExpansion
\frac{1}{4}\phi_{mn}(x,y)\exp i(\theta_{mn}-E_{mn}t) \label{psit}%
\end{equation}
Note that $\psi$ is periodic in time, with period $4\pi$ (since $4\pi E_{mn}$
is always an integer multiple of $2\pi$).

From (\ref{deB}), the velocity components of the particle at any moment are
given by%
\begin{equation}
\frac{dx}{dt}=\frac{i}{2|\psi|^{2}}\left(  \psi\frac{\partial\psi^{\ast}%
}{\partial x}-\psi^{\ast}\frac{\partial\psi}{\partial x}\right)  \label{vx}%
\end{equation}
and%
\begin{equation}
\frac{dy}{dt}=\frac{i}{2|\psi|^{2}}\left(  \psi\frac{\partial\psi^{\ast}%
}{\partial y}-\psi^{\ast}\frac{\partial\psi}{\partial y}\right)  \label{vy}%
\end{equation}

\section{Character of the Trajectories}

The velocities (\ref{vx}) and (\ref{vy}) are ill-defined at nodes (where
$|\psi|=0$), and tend to diverge as nodes are approached. This is because,
close to a node, small displacements in $x$ and $y$ can generate large changes
in phase $S=\operatorname{Im}\ln\psi$, corresponding to a large gradient
$\nabla S$.\footnote{Because $\psi$ is a smooth, single-valued function, a
small displacement $(\delta x,\delta y)$ produces a small change $\delta\psi$
in the complex plane. However, close to a node ($|\psi|=0$), $\delta\psi$ lies
near the origin of the complex plane and so can correspond to a large change
$\delta S$ of phase.} Note, however, that as shown by Berndl \textit{et al}.
\cite{Berndl1,Berndl2}, for reasonable Hamiltonians and wave functions the set
of initial particle positions that reach singularities of the velocity field
in finite time is of $|\psi|^{2}$-measure zero.

Thus, for almost all initial values $(x(0),y(0))$, the trajectory
$(x(t),y(t))$ of the particle may be calculated by numerical integration of
(\ref{vx}) and (\ref{vy}).

\begin{figure}[h!]
  \begin{center}
    \leavevmode
\resizebox{12cm}{10cm}{
        
    \includegraphics{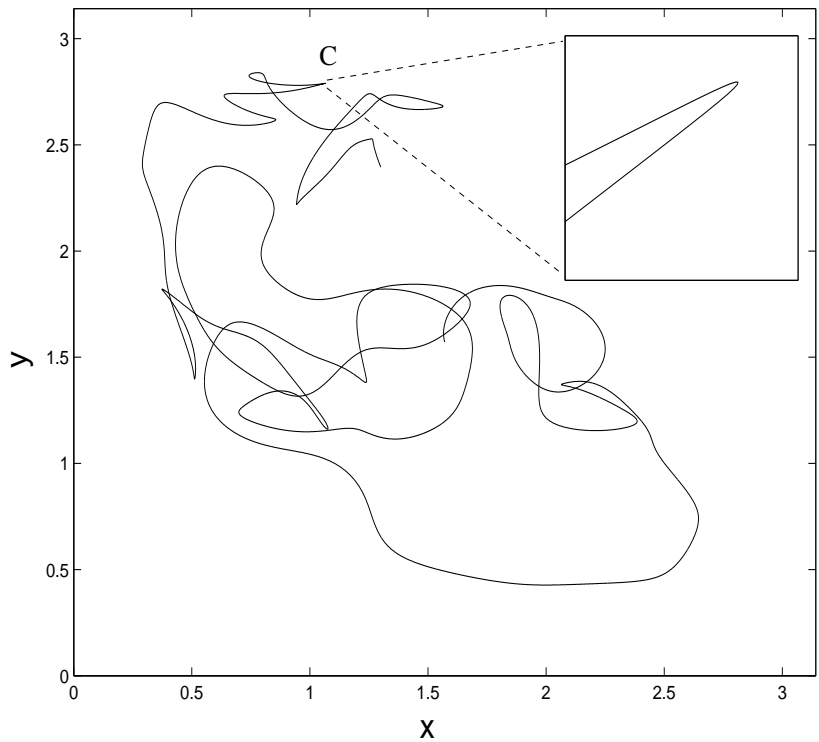}}
    \caption{A typical trajectory, with an apparent cusp at C. The
particle actually turns around very slowly at C.}
  \end{center}
\end{figure}

Our method is based on the Runge-Kutta-Fehlberg algorithm \cite{Recipes}, with
an adaptive time step $h$ adjusted so that the absolute errors in $x(t)$ and
$y(t)$ at each step are both less than $h\Delta$ for some fixed $\Delta$. For
each trajectory, we begin by setting $\Delta=10^{-6}$. We then recalculate the
same trajectory with $\Delta=10^{-7}$. If the global error (the distance
between the final positions) is less than $0.01$, we retain the more accurate
result. Otherwise, we recalculate with $\Delta=10^{-8}$, and so on.

For a small number of trajectories, an accurate calculation was beyond the
bounds of our computational resources, which we fixed at a minimum of
$\Delta=10^{-12}$ and a maximum of $10^{8}$ time steps (per trajectory). Over
a time $2\pi$ the number of problematic trajectories is about 1 in 60,000,
while over a time $4\pi$ it is about 1 in 1,500 (sampling uniformly over the box).

A typical trajectory is shown in Fig. 2. In general, the trajectories are
rather irregular.

At some points, for example at C (Fig. 2), there is an apparent cusp, but
closer examination shows that the tangent to the curve is not discontinuous.
Further, at least in this case, the velocity field at C is in fact very small:
the particle is slowly turning around, with a speed in the range $\sim
10^{-2}-10^{-3}$.

\begin{figure}[h!]
  \begin{center}
    \leavevmode
\resizebox{12cm}{10cm}{
        
    \includegraphics{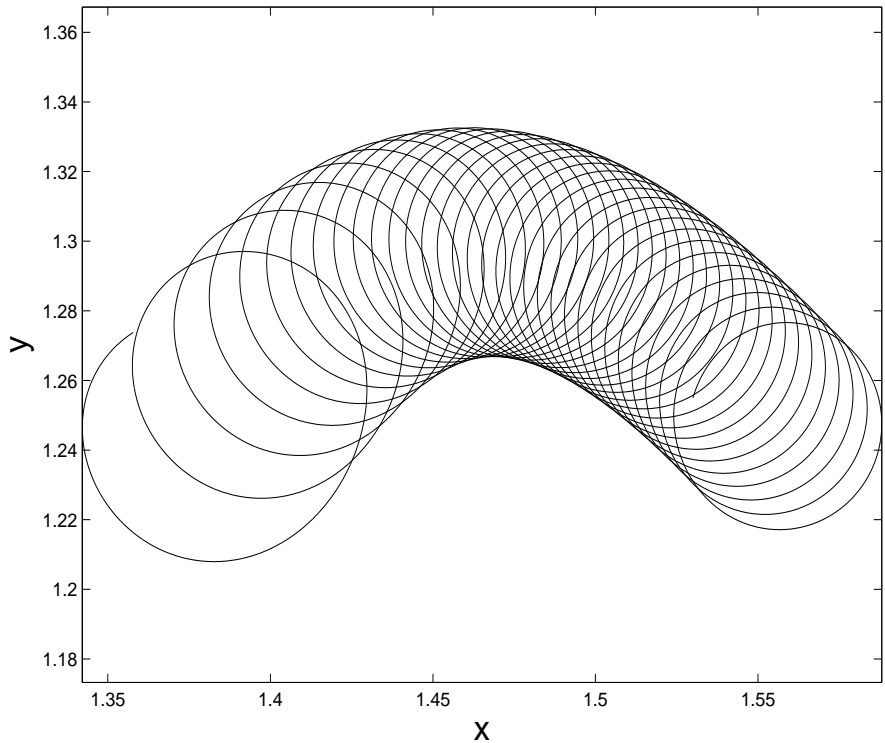}}
    \caption{Close-up of a trajectory near a node or quasi-node (where
$|\psi|$ is known to be very small and possibly zero). The particle rapidly
circles around a moving point at which $1/|\psi|$ is highly peaked.}
  \end{center}
\end{figure}

The motion tends to be particularly rapid in regions where $|\psi|$ is small.
Fig. 3 shows a close-up of a trajectory near a nodal or quasi-nodal point,
where $|\psi|$ is known to be very small (but not known to be strictly zero,
it being impossible to tell in a numerical analysis). The spatial region shown
is about $0.3\%$ of the area of the whole box, and the displayed trajectory
covers a time interval $\Delta t=0.23$.\footnote{The region shown is
$x\in(1.35,1.6)$, $y\in(1.21,1.33)$, and the time interval is $t\in
(9.39,9.62).$} The particle follows a rapid circular motion around a point
moving from right to left in the figure -- and the moving point is a node or
quasi-node, at which $1/|\psi|$ is highly peaked. For the case shown,
$1/|\psi|$ has an apparent peak value of about $60$, and the speed of the
particle varies widely from $\sim10$ to $\sim50$.

\begin{figure}[h!]
  \begin{center}
    \leavevmode
\resizebox{12cm}{10cm}{
        
    \includegraphics{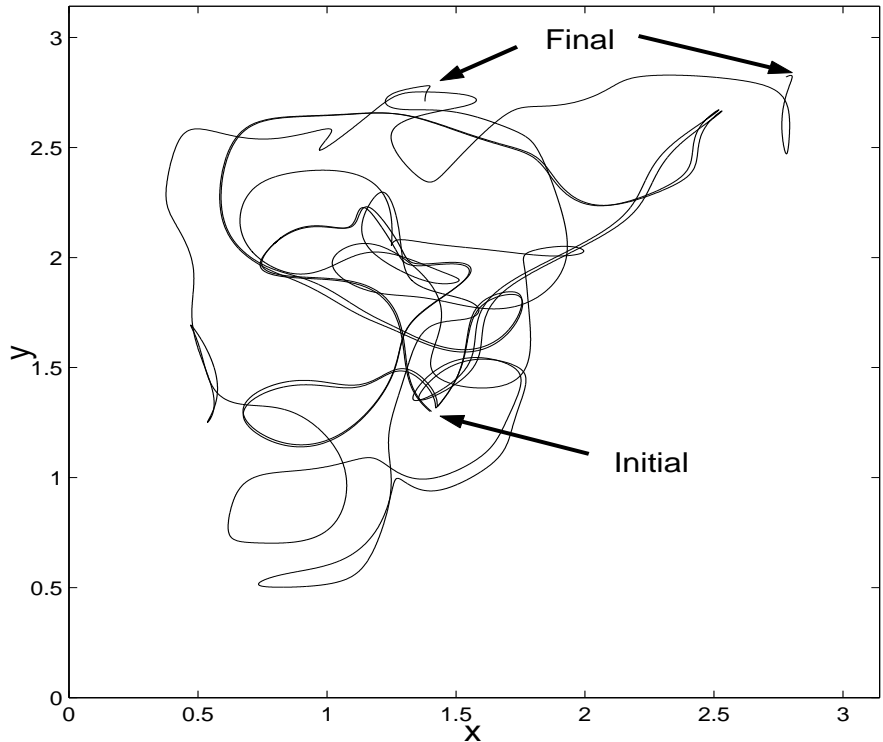}}
    \caption{Illustration of the divergence of neighbouring trajectories.}
  \end{center}
\end{figure}

In Fig. 4, we illustrate the divergence of neighbouring trajectories. Two
distinct but nearby initial positions at $t=0$ evolve after time $t=4\pi$ into
widely-separated final positions. For the example shown, the initial distance
between the points is $0.005$, while the final distance is $1.403$.

The divergence of neighbouring trajectories and the behaviour of Lyapunov
exponents in pilot-wave dynamics has been studied by a number of authors
\cite{PV95,PV96,SF95a,Frisk}. Chaotic trajectories have been reported for a
range of systems: the two-dimensional anisotropic harmonic oscillator
\cite{PV95,PV96}, the quantum kicked rotor \cite{SF95a,DM96}, the quantum cat
map \cite{FS95b}, the hydrogen atom in an external field \cite{Iac96}, the
H\'{e}non-Heiles oscillator \cite{Sen96}, and a two-particle stationary state
\cite{PV97}.

For the two-dimensional square box, chaotic trajectories have been reported by
Frisk \cite{Frisk}, who emphasises the importance of nodes in generating
chaotic motion, and whose numerical simulations suggest a proportionality
between the Lyapunov exponent and the number of nodes.

\section{Relaxation to Quantum Equilibrium}

We now turn to the time evolution of an initial nonequilibrium ensemble. We
assume that every particle in the ensemble is guided by the same wave function
$\psi(x,y,t)$, given by (\ref{psit}). At $t=0$, the (nonequilibrium)
distribution is chosen to be%
\begin{equation}
\rho(x,y,0)=\left(  \frac{2}{\pi}\right)  ^{2}\sin^{2}x\sin^{2}y \label{rho0}%
\end{equation}
This is just $|\phi_{11}(x,y)|^{2}$ -- that is, the ground-state equilibrium
distribution. It is displayed in Fig. 5.

\begin{figure}[h!]
  \begin{center}
    \leavevmode
\resizebox{12cm}{10cm}{
        
    \includegraphics{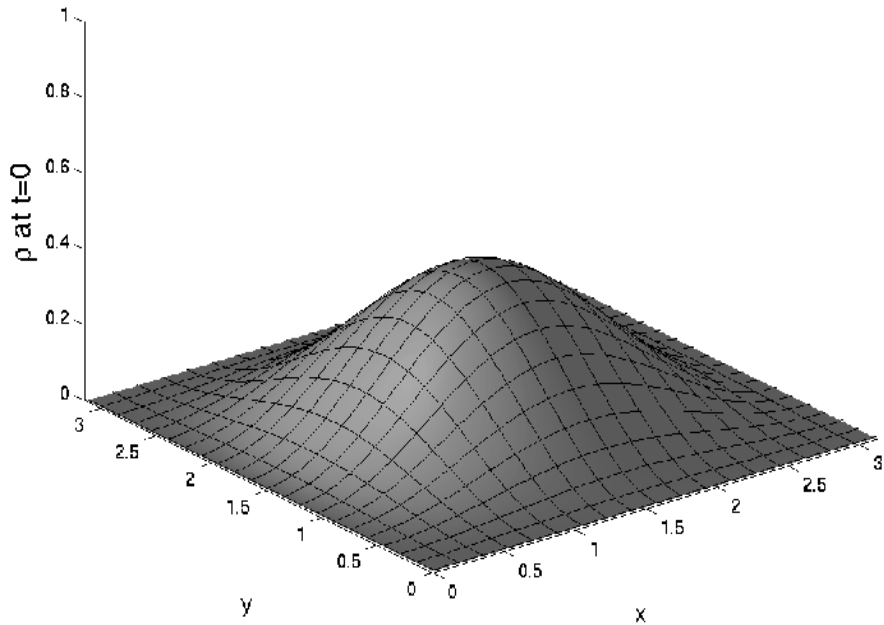}}
    \caption{The initial nonequilibrium distribution, chosen to be
$\rho(x,y,0)=|\phi_{11}(x,y)|^{2}$ (equal to the ground-state equilibrium distribution).}
  \end{center}
\end{figure}

Instead of integrating the continuity equation (\ref{Cont}) directly, it is
more convenient to calculate the trajectories, from which it is
straightforward to deduce the evolution of any initial distribution
$\rho(x,y,0)$.

From (\ref{Cont}) and (\ref{Contpsi2}), it follows that the ratio%
\begin{equation}
f(x,y,t)\equiv\frac{\rho(x,y,t)}{|\psi(x,y,t)|^{2}}%
\end{equation}
is conserved along trajectories: $df/dt=0$ or%
\[
\frac{\partial f}{\partial t}+\frac{dx}{dt}\frac{\partial f}{\partial x}%
+\frac{dy}{dt}\frac{\partial f}{\partial y}=0
\]
If the initial point $(x_{0},y_{0})$ evolves into $(x,y)$ at time $t$, then
for any ensemble%
\begin{equation}
f(x,y,t)=f(x_{0},y_{0},0)
\end{equation}

Thus, given the function%
\[
f(x_{0},y_{0},0)=\frac{\rho(x_{0},y_{0},0)}{|\psi(x_{0},y_{0},0)|^{2}}%
\]
at $t=0$, the distribution $\rho(x,y,t)$ at later times may be written as%
\begin{equation}
\rho(x,y,t)=|\psi(x,y,t)|^{2}f(x_{0},y_{0},0) \label{rhodeduce}%
\end{equation}

Given the mapping from $(x_{0},y_{0})$ to $(x,y)$, the fate of any initial
distribution $\rho(x,y,0)$ may be deduced from (\ref{rhodeduce}).

One might begin with a uniform lattice of initial points $(x_{0},y_{0})$, and
calculate the future trajectories $(x(t),y(t))$, using (\ref{rhodeduce}) to
deduce the distribution $\rho(x(t),y(t),t)$ at the later points $(x(t),y(t))$.
This method was used in the first relaxation simulation, for the simple case
of a one-dimensional box \cite{AVth}. However, the lattice of points at later
times is unfortunately distorted, and there appear large regions containing
hardly any lattice points at all.

A more accurate procedure is to evolve trajectories \textit{backwards}, from
$(x,y)$ at any desired $t$ to the starting point $(x_{0},y_{0})$ at $t=0$
\cite{HW}. Thus, to calculate $\rho(x,y,t)$ at time $t$, we set up a uniform
lattice of points $(x,y)$ at time $t$ and backtrack the trajectories to the
initial points $(x_{0},y_{0})$ at $t=0$. Knowing the function $f(x_{0}%
,y_{0},0)$, we may then use (\ref{rhodeduce}) to deduce $\rho(x,y,t)$.

This backtracking procedure has the disadvantage that one has to calculate the
trajectories over the whole interval from $t$ to $t=0$, for every desired
value of $t$ (for which one wishes to know $\rho(x,y,t)$). But overall the
procedure is better, because it generates results for $\rho(x,y,t)$ on a
uniform lattice. This is particularly important when one wishes to calculate
the coarse-grained \textit{H}-function.

As mentioned in Sect. 3, for some trajectories an accurate calculation is
beyond our computational resources ($\Delta=10^{-12}$ and $10^{8}$ time steps
per trajectory). Nevertheless, for practical reasons of data handling, it is
convenient to assign values to all of them, rather than ignore the problematic
cases.\footnote{Some cases required more than $10^{8}$ time steps even for
$\Delta=10^{-6}$; each of these were aborted and assigned a value from a
neighbouring trajectory. In other cases, decreasing $\Delta$ exceeded the
limit of $10^{8}$ time steps, and we retained the value obtained using the
smallest value of $\Delta$ that did not exceed this limit (despite the global
error exceeding $0.01$). Finally, some cases reached the minimum setting
$\Delta=10^{-12}$ without achieving the desired limit on the global error, and
for these we assigned the value obtained using $\Delta=10^{-12}$.} Since the
number of problematic trajectories is found to be small (1 in 60,000 over a
time $2\pi$ and 1 in 1,500 over a time $4\pi$), they have a negligible effect
on the results reported below.

Note that because the values of $f=\rho/|\psi|^{2}$ are carried along
trajectories, the exact (fine-grained) value of $\rho(x,y,t)$ will always
differ from $|\psi(x,y,t)|^{2}$, by just the same multiplicative factor
$f(x_{0},y_{0},0)$ by which $\rho(x_{0},y_{0},0)$ differed from $|\psi
(x_{0},y_{0},0)|^{2}$. Indeed, the property $df/dt=0$ is analogous to the
Liouville property in classical statistical mechanics, according to which the
phase-space density is constant along Hamiltonian trajectories (see Sect. 5
below). In both cases, relaxation occurs only in a coarse-grained sense.

\begin{figure}[h!]
  \begin{center}
    \leavevmode
\resizebox{12cm}{10cm}{
        
    \includegraphics{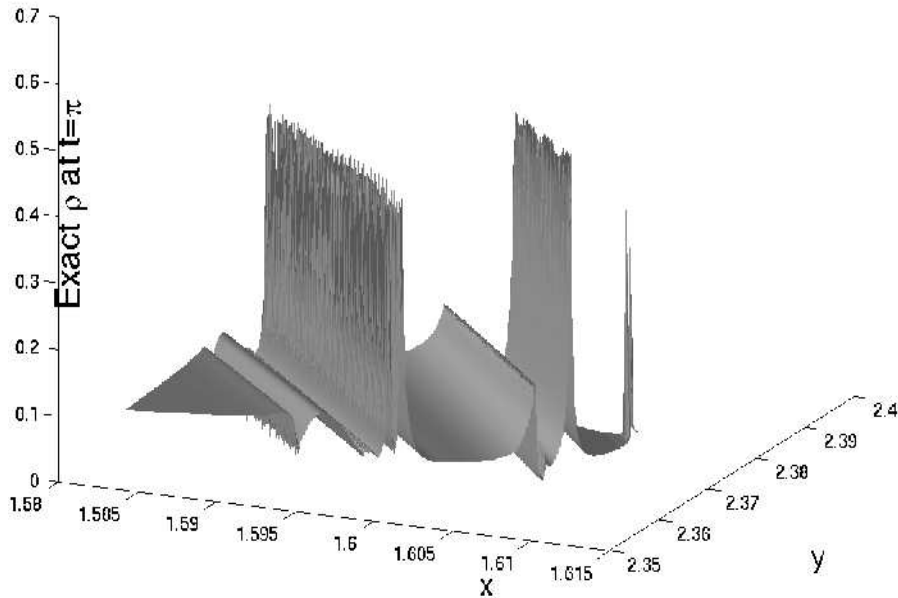}}
    \caption{Close-up of the exact $\rho(x,y,t)$ in a small spatial
region, covering just $0.006\%$ of the area of the box, at time $t=\pi$. The
fine-grained distribution is very irregular.}
  \end{center}
\end{figure}

To see how this works, in Fig. 6 we show a close-up of $\rho(x,y,t)$ in a
small fixed region of the box, at time $t=\pi$. The exact, fine-grained
distribution is highly irregular. The area shown is very small: it has side
$\pi/128$, covering just $0.006\%$ of the total area of the box. (For this
calculation, we used a uniform lattice of $40,000$ points in this small area,
with the lattice specified at $t=\pi$.)

Clearly, the exact fine-grained distribution is extremely complex, and it is
useful to consider its appearance under coarse-graining. Given a (square)
coarse-graining cell of side $\varepsilon$, the distribution $\rho$ may be
averaged over the cell to produce a coarse-grained value%
\begin{equation}
\frac{1}{\varepsilon^{2}}\int\int_{\mathrm{cell}}dxdy\ \rho\label{cging}%
\end{equation}
which may be assigned to the centre of the cell.

If the cells do not overlap, we denote the coarse-grained distribution by
$\bar{\rho}$. Results for $\bar{\rho}$ will be shown in Sect. 5 below (using a
coarse-graining length $\varepsilon=\pi/32$).

A smoother coarse-grained distribution, denoted $\tilde{\rho}$, is obtained if
one takes overlapping cells (a procedure used in quantum chemistry, for
example). To generate $\tilde{\rho}$, we take cells of side $\varepsilon
=\pi/16$ that overlap with their immediate neighbours in $88\%$ of their area.
Specifically, given one cell, shifting it along $x$ or $y$ by a distance equal
to $12\%$ of its length (that is, $12\%$ of $\pi/16$) generates the
neighbouring cell, and so on. In this way, in the box of area $\pi^{2}$ we
obtain a patchwork of $126\times126$ overlapping cells, each with its own
value of $\tilde{\rho}$. For the trajectory calculations, we use a lattice of
$400\times400$ particles, with $25\times25$ particles in each cell.

\begin{figure}[h!]
  \begin{center}
    \leavevmode
\resizebox{15cm}{15cm}{
        
    \includegraphics{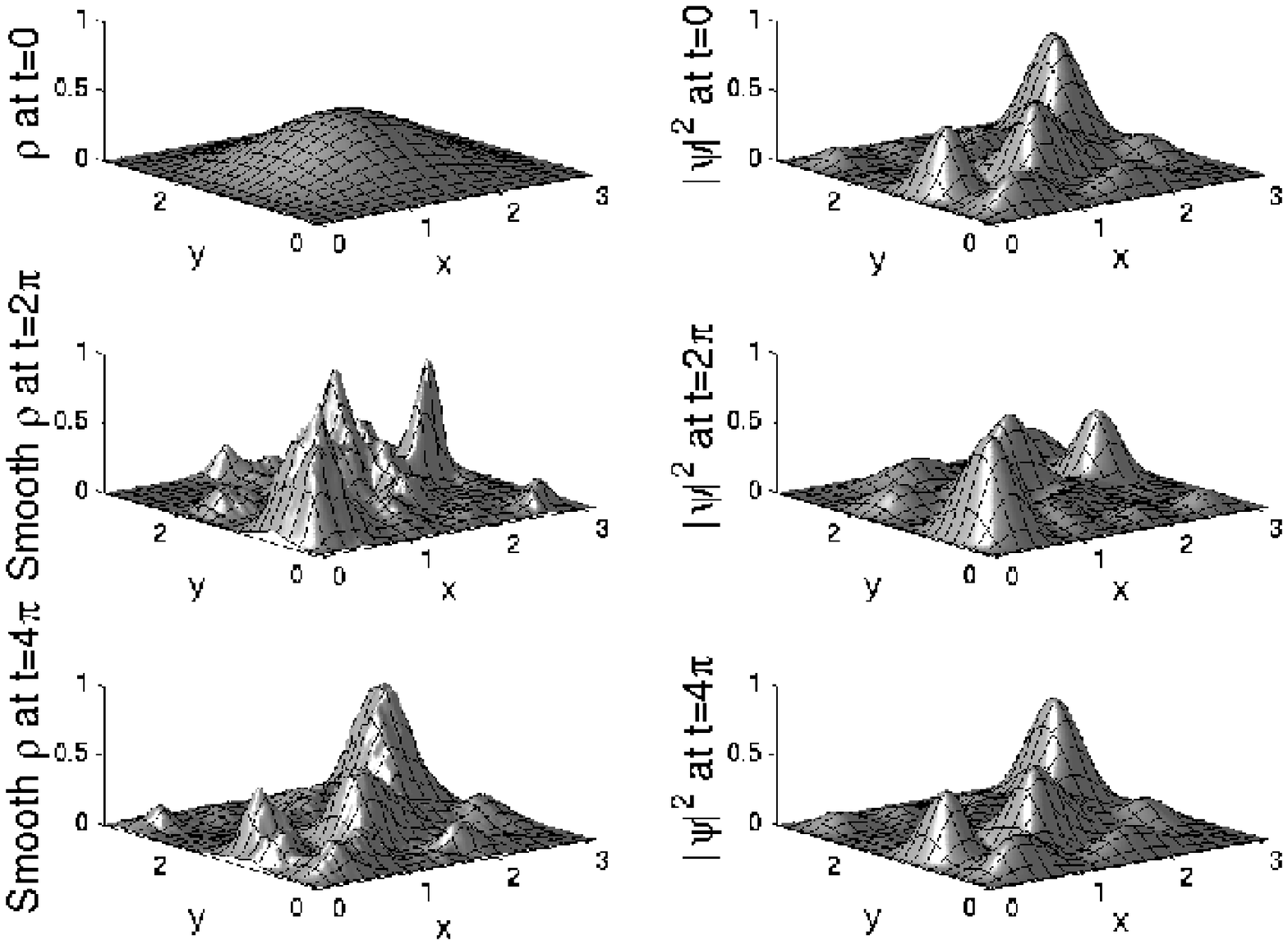}}
    \caption{Smoothed $\tilde{\rho}$, compared with $|\psi|^{2}$, at times
$t=0$, $2\pi$ and $4\pi$. While $|\psi|^{2}$ recurs to its initial value, the
smoothed $\tilde{\rho}$ shows a remarkable evolution towards equilibrium.}
  \end{center}
\end{figure}

We have calculated the evolution of $\tilde{\rho}$ up to $t=4\pi$ (when
$|\psi|^{2}$ recurs to its initial value). This is shown in Fig. 7. Displayed
are $\tilde{\rho}$ and $|\psi|^{2}$, at times $t=0$, $2\pi$ and $4\pi$.
Evidently, $\tilde{\rho}$ and $|\psi|^{2}$ become rather close, and a
remarkable relaxation towards quantum equilibrium has occurred after just one
time cycle of $4\pi$.

Fig. 8 shows the same data, as contour plots.

\begin{figure}[h!]
  \begin{center}
    \leavevmode
\resizebox{15cm}{15cm}{
        
    \includegraphics{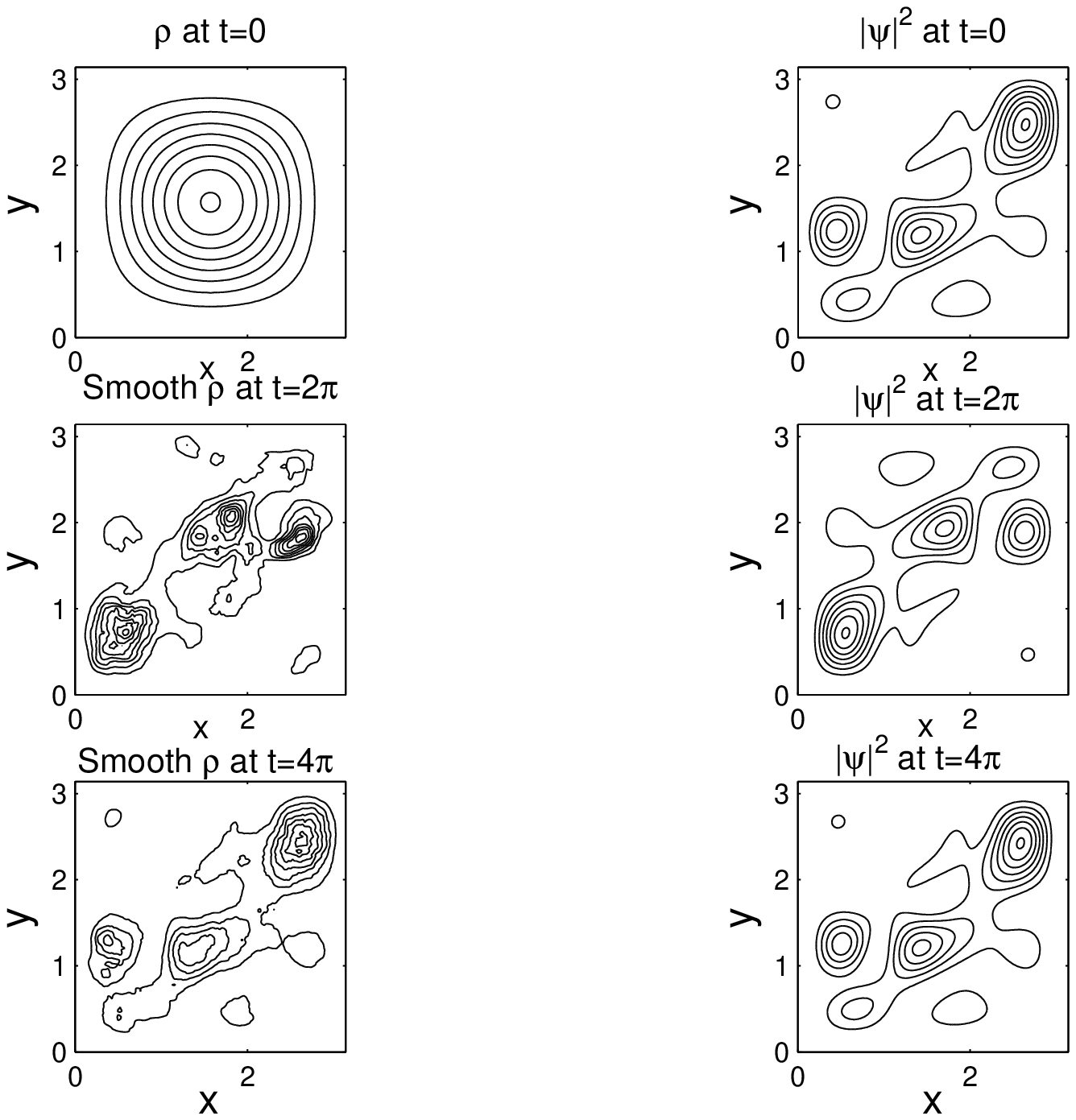}}
    \caption{Smoothed $\tilde{\rho}$, compared with $|\psi|^{2}$, at times
$t=0$, $2\pi$ and $4\pi$. The same data as in Fig. 8, displayed as contour plots.}
  \end{center}
\end{figure}

Note that while $\psi$ and its associated velocity field are periodic, in
general the trajectories are not, so that the distribution $\rho$ does
\textit{not} recur at $t=4\pi$. This is in contrast with the one-dimensional
case, where the particles cannot move past each other, constraining each
trajectory to recur simultaneously with $|\psi|^{2}$, so that any initial
$\rho_{0}\neq|\psi_{0}|^{2}$ recurs as well \cite{AVth,ValIsch}.

Presumably, evolution over more time cycles of $4\pi$ will make $\tilde{\rho}$
converge ever closer to $|\psi|^{2}$. However, we have not extended our
calculations beyond $t=4\pi$. Possibly, a small residual nonequilibrium will
remain after an arbitrary number of cycles, but this seems unlikely.

Given the mapping from $(x_{0},y_{0})$ to $(x(4\pi),y(4\pi))$, not only can
one immediately construct $\rho(x,y,4\pi)$ from any initial $\rho(x,y,0)$, one
can also extend the evolution to an arbitrary number of cycles $4\pi$ -- that
is, to times $t=4\pi n$ for positive integral $n$ -- by simple iteration of
the mapping (the velocity field being periodic with period $4\pi$). It might
then appear that, given the calculations performed so far, we could easily
extend them to $t=4\pi n$. Unfortunately, the calculated map from
$(x_{0},y_{0})$ to $(x(4\pi),y(4\pi))$ cannot be reapplied to map
$(x(4\pi),y(4\pi))$ to $(x(8\pi),y(8\pi))$ and so on, because the backtracking
method we have used distorts the lattice at $t=0$.

\section{Decay of the \textit{H}-Function; Relaxation Timescale}

The difference between $\rho$ and $|\psi|^{2}$ may be quantified by the value
of the \textit{H}-function (\ref{Hfn}). As we mentioned in the Introduction,
this is just minus the relative entropy of $\rho$ with respect to $|\psi|^{2}$
(a standard measure of the difference between two distributions). The
\textit{H}-function is bounded below by zero, and is equal to zero if and only
if $\rho=|\psi|^{2}$ everywhere \cite{PLA1,AVth,ValIsch}.

Note, however, that because of (\ref{Cont}) and (\ref{Contpsi2}) the exact $H$
is constant in time, $dH/dt=0$, reflecting the fact (already mentioned) that
differences between fine-grained values of $\rho$ and $|\psi|^{2}$ never
disappear (the ratio $f=\rho/|\psi|^{2}$ being carried along trajectories). A
decreasing $H$ is obtained only under coarse-graining.

The situation is similar classically, where departures from thermal
equilibrium may be quantified in terms of the classical \textit{H}-function
(on phase space)%
\[
H_{\mathrm{class}}=\int\int dqdp\;\rho\ln\rho
\]
which is just minus the relative entropy of $\rho$ with respect to the uniform
distribution. Under Hamiltonian evolution, $d\rho/dt=0$ along trajectories
(Liouville's theorem), and the exact $H_{\mathrm{class}}$ is constant,
$dH_{\mathrm{class}}/dt=0$. However, in appropriate circumstances, the
classical phase-space evolution defined by (\ref{Liou}) leads to thermal
relaxation on a coarse-grained level. This may be quantified in terms of the
classical coarse-grained \textit{H}-function%
\[
\bar{H}_{\mathrm{class}}=\int\int dqdp\;\bar{\rho}\ln\bar{\rho}%
\]
where $\bar{\rho}$ is $\rho$ averaged over (non-overlapping) coarse-graining
cells. This obeys the coarse-graining \textit{H}-theorem \cite{Tol,Dav}%
\[
\bar{H}_{\mathrm{class}}(t)\leq\bar{H}_{\mathrm{class}}(0)
\]
(assuming no initial fine-grained microstructure in $\rho$ at $t=0$), and
$\bar{H}_{\mathrm{class}}$ is minimised by $\bar{\rho}=\mathrm{const}.$. This
theorem formalises the intuitive idea of Gibbs -- that an initial non-uniform
distribution will tend to develop fine-grained structure and become more
uniform on a coarse-grained level.

In the pilot-wave case too, the evolution (on configuration space) may lead to
relaxation on a coarse-grained level, as shown by the above numerical
simulations. And the relaxation may be similarly quantified in terms of the
coarse-grained \textit{H}-function%
\begin{equation}
\bar{H}=\int dq\;\bar{\rho}\ln(\bar{\rho}/\overline{\left\vert \psi\right\vert
^{2}}) \label{Hbar}%
\end{equation}
which obeys the \textit{H}-theorem \cite{PLA1,AVth,ValIsch}%
\[
\bar{H}(t)\leq\bar{H}(0)
\]
(assuming no initial fine-grained microstructure in $\rho$ and $\left\vert
\psi\right\vert ^{2}$), and where again $\bar{H}\geq0$ for all $\bar{\rho}$,
$\overline{\left\vert \psi\right\vert ^{2}}$ and $\bar{H}=0$ if and only if
$\bar{\rho}=\overline{\left\vert \psi\right\vert ^{2}}$ everywhere. This
version of the \textit{H}-theorem formalises the idea that $\rho$ and
$\left\vert \psi\right\vert ^{2}$ behave like two `fluids' which are `stirred'
by the same velocity field (since they obey the same continuity equation), so
that $\rho$ and $\left\vert \psi\right\vert ^{2}$ tend to become
indistinguishable on a coarse-grained level.\footnote{Note, however, that the
velocity field (\ref{deB}) is related to $\psi$ by the Schr\"{o}dinger
equation, as well as by the continuity equation, whereas no such additional
relation exists between the velocity field and $\rho$. This explains why
$\rho$ develops a filamentary structure while $|\psi|^{2}$ does not
\cite{AVth}.}

A natural timescale $\tau$ for relaxation may be defined in terms of the time
derivatives of $\bar{H}(t)$ at $t=0$.

Using the continuity equations (\ref{Cont}) and (\ref{Contpsi2}), it follows
that $\left(  d\bar{H}/dt\right)  _{0}=0$ and (for $q=(x,y)$) \cite{AVth}%
\begin{equation}
\left(  \frac{d^{2}\bar{H}}{dt^{2}}\right)  _{0}=-\int\int dxdy\frac{|\psi
_{0}|^{2}}{f_{0}}\left(  \overline{(\dot{X}_{0}\cdot\nabla f_{0})^{2}%
}-\overline{(\dot{X}_{0}\cdot\nabla f_{0})}^{2}\right)  \leq0\label{H2dots}%
\end{equation}
where $\psi_{0}=\psi(x,y,0)$, $f_{0}=f(x,y,0)$, $\dot{X}_{0}\equiv(\dot{x}%
_{0},\dot{y}_{0})$ and $\nabla\equiv(\partial/\partial x,\partial/\partial
y)$. (The quantity in brackets is non-negative, and is equal to the spatial
variance of $\dot{X}_{0}\cdot\nabla f_{0}$ over a coarse-graining cell.)

Thus we may define $\tau$ by \cite{AVth}%
\begin{equation}
\frac{1}{\tau^{2}}\equiv-\frac{\left(  d^{2}\bar{H}/dt^{2}\right)  _{0}}%
{\bar{H}_{0}}%
\end{equation}
where (\ref{H2dots}) gives $\left(  d^{2}\bar{H}/dt^{2}\right)  _{0}$ in terms
of the initial conditions $\psi_{0}$ and $\rho_{0}$.

Expanding $\dot{X}_{0}\cdot\nabla f_{0}$ in a Taylor series within each
coarse-graining cell of area $\varepsilon^{2}$, it is found that%
\begin{equation}
\left(  d^{2}\bar{H}/dt^{2}\right)  _{0}=-\frac{\varepsilon^{2}}{12}I+O\left(
\varepsilon^{4}\right)
\end{equation}
where \cite{ValIsch,AVbook}%
\begin{equation}
I\equiv\int\int dxdy\ \frac{|\psi_{0}|^{2}}{f_{0}}|\nabla(\dot{X}_{0}%
\cdot\nabla f_{0})|^{2}%
\end{equation}

Thus%
\begin{equation}
\tau=\frac{1}{\varepsilon}\sqrt{\frac{12\bar{H}_{0}}{I}}+O(\varepsilon)
\end{equation}

If $\varepsilon$ is small with respect to the lengthscale over which $\dot
{X}_{0}\cdot\nabla f_{0}$ varies, then $\tau\propto1/\varepsilon$. Taking
$\bar{H}_{0}\sim1$ (a mild nonequilibrium), a rough estimate of $I$ yields the
quoted result (\ref{tau}), in terms of the quantum energy spread $\Delta E$ of
$\psi_{0}$ \cite{ValIsch,AVbook}.

Note that, given the result $\tau\propto1/\varepsilon$ (for small
$\varepsilon$), the formula (\ref{tau}) may be obtained on dimensional
grounds. And in any case, (\ref{tau}) is only a rough, order-of-magnitude
estimate. For our initial wave function (\ref{psi0}), $\Delta E\simeq4$ and
(in our units)%
\begin{equation}
\tau\sim\frac{1}{8\varepsilon} \label{tau2}%
\end{equation}
Thus, at least initially, $\bar{H}=\bar{H}(t)$ should decay on a timescale of
order $1/8\varepsilon$.

To investigate the decay of $\bar{H}(t)$ numerically, we must work in terms of
the coarse-grained quantities $\bar{\rho}$ and $\overline{\left\vert
\psi\right\vert ^{2}}$, with non-overlapping coarse-graining cells (as used in
the above \textit{H}-theorem).

\begin{figure}[h!]
  \begin{center}
    \leavevmode
\resizebox{14cm}{12cm}{
        
    \includegraphics{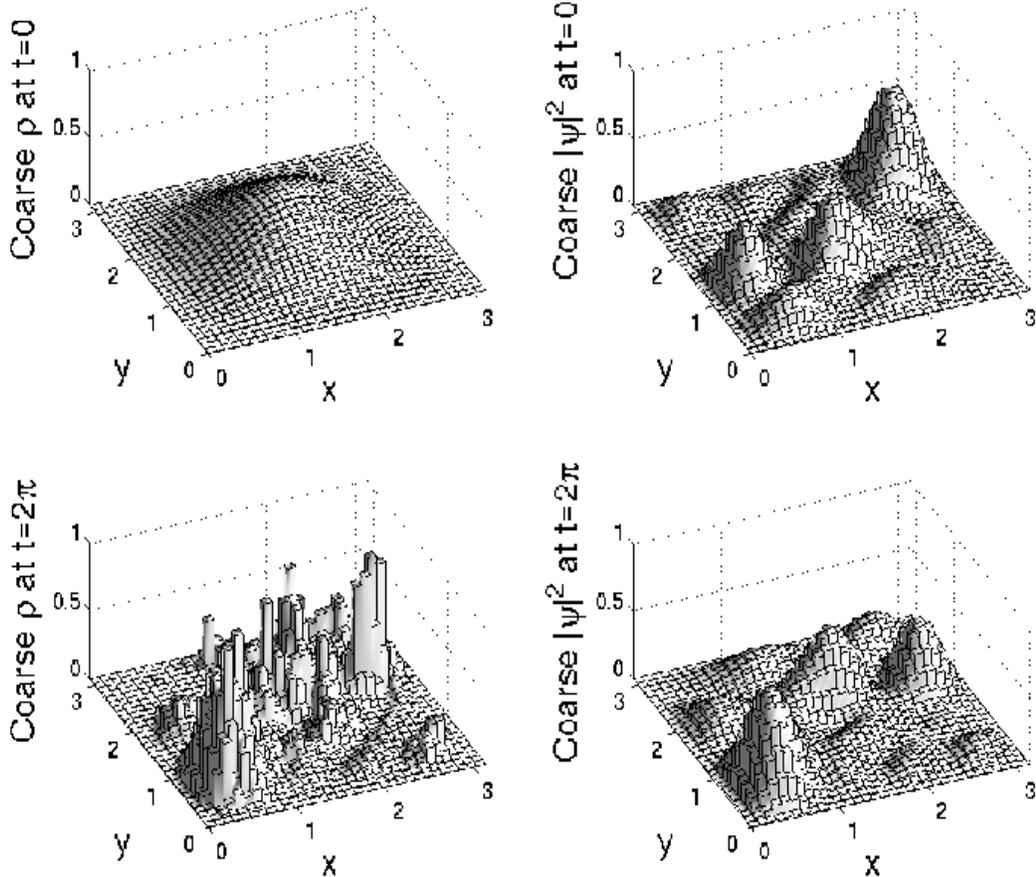}}
    \caption{Coarse-grained $\bar{\rho}$ and coarse-grained $\overline
{\left\vert \psi\right\vert ^{2}}$ (with non-overlapping cells) at times $t=0$
and $t=2\pi$.}
  \end{center}
\end{figure}

Fig. 9 displays $\bar{\rho}$ and $\overline{\left\vert \psi\right\vert ^{2}}$,
with non-overlapping cells of side $\varepsilon=\pi/32$, at times $t=0$ and
$t=2\pi$.

\begin{figure}[h!]
  \begin{center}
    \leavevmode
\resizebox{12cm}{10cm}{
        
    \includegraphics{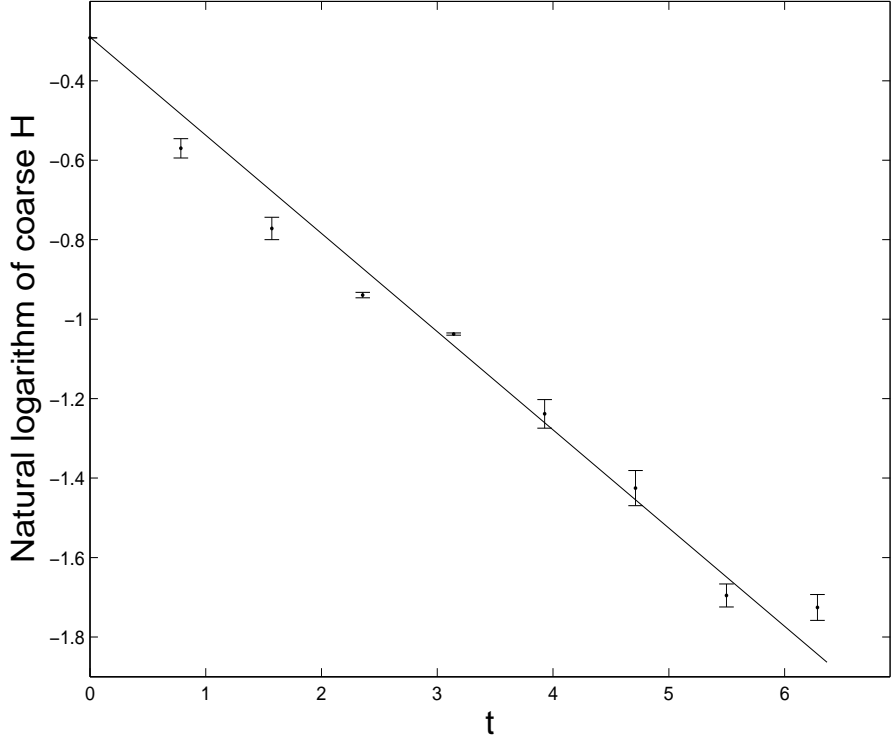}}
    \caption{Plot of $\ln\bar{H}$ against time, from $t=0$ to $t=2\pi$,
showing an approximately exponential decay of $\bar{H}(t)$ with a time
constant $t_{c}\approx4$.}
  \end{center}
\end{figure}

In Fig. 10, we plot $\ln\bar{H}$ against time from $t=0$ to $t=2\pi$ (where
the value of $\bar{H}$ was calculated every $\pi/4$ time units). The evolution
of $\ln\bar{H}$ is approximately linear, corresponding to an approximately
exponential decay $\bar{H}(t)\approx\bar{H}_{0}e^{-t/t_{c}}$, with a
(best-fitting) time constant $t_{c}\approx4$. This compares favourably with
the order-of-magnitude estimate (\ref{tau2}), which for the present
coarse-graining length $\varepsilon=\pi/32$ gives $\tau\sim1.3$.

Fig. 10 shows error bars for the displayed values of $\ln\bar{H}$. These were
obtained as follows. The main source of error is the approximation of
$\bar{\rho}$ by an average over a finite number of points within a
coarse-graining cell. To estimate this error, each calculation of $\bar{\rho}$
is done again with a different lattice of points, corresponding to a different
sampling within the cells (first with a lattice of $25\times25$ particles,
then with $27\times27$). The two results yield different values for $\bar{H}$;
however, the difference is usually only about $2\%$, and never more than
$4\%$. Another (considerably smaller) source of error in $\bar{H}$ is of
course the inaccuracy in the calculated particle positions. This error is
estimated by comparing values of $\bar{H}$ obtained with local errors $\Delta$
(in the trajectories) differing by a factor of ten; the difference in values
of $\bar{H}$ is found to be quite negligible.

Note that the natural quantum timescale for this system is $\Delta t\sim
\hbar/\Delta E\sim0.25$, which is arguably not much further from $t_{c}$ than
is $\tau$. However, generally speaking, the quantum timescale $\Delta
t\sim\hbar/\Delta E$ for a system is unrelated to the estimated relaxation
timescale $\tau\sim(1/\varepsilon)\hbar^{2}/m^{1/2}(\Delta E)^{3/2}$ in
(\ref{tau}), since $\Delta t$ has no dependence on $\varepsilon$ and scales
differently with $\Delta E$. Writing $\Delta E\sim(\Delta p)^{2}/2m$, the
special choice of $\varepsilon\sim\hbar/\Delta p$ for the coarse-graining
length leads to $\tau\sim\Delta t$, and our estimated $\tau$ is not far from
$\Delta t$ only because our $\varepsilon$ is not far from $\hbar/\Delta p$.

It would be interesting to check numerically the dependence of the relaxation
timescale on the coarse-graining length $\varepsilon$ and on the energy spread
$\Delta E$. This has been done for the rather artificial but much simpler case
of the one-dimensional box (in which trajectories cannot move past each
other), with results that accord quite well with the predicted scalings
$\tau\propto1/\varepsilon$ and $\tau\propto1/(\Delta E)^{3/2}$
\cite{ValIsch,DSV}. To perform the corresponding checks for the
two-dimensional case is a task for the future.

We have found an approximately exponential decay of $\bar{H}$ with time.
Presumably this behaviour could be derived, by supplementing the underlying
dynamics with some sort of phenomenological Markovian assumption, analogous to
the classical hypothesis of molecular chaos at every instant (from which the
Boltzmann equation may be derived).

\section{Discussion and Conclusion}

We have seen that the simple choice of an initial nonequilibrium state
(\ref{rho0}) relaxes towards equilibrium rather efficiently, with an
accompanying (approximately exponential) decrease in the coarse-grained
\textit{H}-function on the expected timescale.

Of course, this is not to imply that \textit{any} initial distribution
$\rho_{0}\neq|\psi_{0}|^{2}$ would relax to equilibrium: that is impossible in
any time-reversal invariant theory. Just as in classical mechanics, for every
solution of the dynamical laws evolving towards equilibrium, there is a
time-reversed solution evolving away from equilibrium. For example, as new
initial conditions $\psi_{0}^{\prime}$, $\rho_{0}^{\prime}$ at $t=0$ we could
take%
\begin{equation}
\psi^{\prime}(x,y,0)=\psi^{\ast}(x,y,\pi),\;\;\;\rho^{\prime}(x,y,0)=\rho
(x,y,\pi) \label{r}%
\end{equation}
where $\rho(x,y,\pi)$ is the former (exact) distribution at $t=\pi$, whose
complex fine-grained microstructure is shown in close-up in Fig. 6. The
dynamical equations (\ref{Sch}) and (\ref{Cont}) imply that these conditions
will evolve into%
\begin{equation}
\psi^{\prime}(x,y,t)=\psi^{\ast}(x,y,\pi-t),\;\;\;\rho^{\prime}(x,y,t)=\rho
(x,y,\pi-t)
\end{equation}
so that after a time $t=\pi$ we recover our former initial distribution
$\rho(x,y,0)$ given by (\ref{rho0}) -- which is further away from equilibrium
than the present initial state (\ref{r}). Thus the initial conditions
(\ref{r}) evolve away from equilibrium, and the coarse-grained \textit{H}%
-function increases with time.

However, as in ordinary statistical mechanics, relaxation towards equilibrium
will be obtained if one assumes that the initial state satisfies certain
conditions -- such as the absence of fine-grained microstructure -- which are
violated by time-reversed `initial' states such as (\ref{r}). Conceptually,
the situation here is the same as in ordinary statistical mechanics.

Conceptual issues in statistical mechanics arguably reduce to issues about
initial conditions, and these are inevitably bound up with questions of
cosmology \cite{Dav,Sklar,Hall,Savitt,ValIsch}. According to our current
understanding, we see thermal nonequilibrium in our universe today only
because gravity has the remarkable property of amplifying small
inhomogeneities in temperature and energy density, leading to the formation of
large-scale structure out of a primordial homogeneous state of thermal
equilibrium \cite{Pad}. In the absence of gravity, a classical isolated system
with no initial fine-grained structure would be expected to evolve towards
thermal equilibrium, just as in pilot-wave dynamics.

Initial conditions can be questioned. A lot of current work in cosmology is
motivated by the desire to explain, for example, why the early universe was so
smooth and homogeneous, and the prevailing view is that the homogeneity arose
from interactions taking place at even earlier times (within cosmological
horizons). Similarly, in the de Broglie-Bohm formulation of quantum theory,
one may ask why all physical systems probed so far follow the Born rule
$\rho=|\psi|^{2}$. Because these systems all have a long and violent
astrophysical history, it is reasonable to explain the distribution
$\rho=|\psi|^{2}$ in terms of a relaxation process from an earlier
nonequilibrium state $\rho\neq|\psi|^{2}$. This leads naturally to the
suggestion that quantum nonequilibrium may have existed in the early universe,
in which case the quantum noise we see today may be regarded as a remnant of
the big bang \cite{PLA2,AVth,AV96,ValIsch,AVbook}. A similar view may be taken
in any deterministic hidden-variables theory \cite{PLAc,Cracow,Sig}.

In the original pilot-wave version of the \textit{H}-theorem \cite{PLA1}, it
was argued that relaxation $P\rightarrow|\Psi|^{2}$ would occur (on a
coarse-grained level) for an ensemble of many-body systems with distribution
$P(\mathbf{x}_{1},\mathbf{x}_{2},....\mathbf{x}_{N},t)$ and wave function
$\Psi(\mathbf{x}_{1},\mathbf{x}_{2},....\mathbf{x}_{N},t)$ (with $N$ large).
It was then shown that single particles extracted from the (eventual)
equilibrium ensemble and prepared with wave function $\psi$ would have a
distribution $\rho=|\psi|^{2}$. However, it is now clear from the above
simulations that a large number of degrees of freedom are not needed for
efficient relaxation to occur. Even for a single particle, relaxation will
occur rapidly if its wave function is a superposition of even a modest number
of energy eigenfunctions, as our numerical example shows.

The simulations performed in this paper add substance to the view, already
mentioned, that relaxation occurs because $\rho$ and $\left\vert
\psi\right\vert ^{2}$ evolve like two `fluids' which are `stirred' by the same
velocity field. The most efficient mixing is found to occur in the
neighbourhood of nodes or quasi-nodes, where $\left\vert \psi\right\vert $ is
small. These points move around inside the box, rather like `electric mixers'
(or stirring devices) moving through a fluid, generating an efficient
relaxation everywhere.

We mentioned in Sect. 2 that the importance of nodes in generating chaotic
motion was noted by Frisk \cite{Frisk}. Frisk also suggested that nodes would
be important if the motion is to have the appropriate mixing properties
required for complete relaxation to quantum equilibrium. The above simulations
certainly bear out the expectation that nodes provide a particularly efficient
mixing. But whether or not relaxation would take place even in their absence
(perhaps on much longer timescales) is not known.

Many details remain to be understood. We are far from a full understanding of
pilot-wave theory as a dynamical system. General properties of the flow, such
as ergodicity and mixing (in a rigorous sense), remain to be investigated.
Still, the results already in hand strongly suggest that, in de Broglie-Bohm
theory, the distribution $\rho=|\psi|^{2}$ is an equilibrium state quite
analogous to thermal equilibrium in classical physics.

It should be noted that the status of the Born rule has been a contentious
issue in quantum theory generally, perhaps most notably in the many-worlds
formulation of Everett \cite{deWG,Deutsch,Wallace}. Some recent authors
\cite{Fuchs1,Fuchs2} base their justification of the Born rule on Gleason's
theorem \cite{G57}, which states that the Born rule is the unique probability
assignment satisfying `noncontextuality' -- the condition that the probability
for an observable should not depend on which other (commuting) observables are
simultaneously measured. However, as pointed out by Bell \cite{Bell66},
Gleason's noncontextuality condition is very strong, as it amounts to assuming
that mutually-incompatible experimental arrangements yield the same statistics
for the observable in question. Saunders \cite{SS} gives an `operational' derivation of the Born rule (in which it is assumed that probabilities are determined by the
quantum state alone); while Zurek \cite{Zu} appeals to `environment-assisted
invariance'. Other recent derivations of the Born rule
arise from novel axioms for quantum theory \cite{LH,Bub}.

Like Euclid's axiom of parallels in geometry, the Born rule seems to stand
apart from the other axioms of quantum theory, and there have been a number of
attempts to derive it either from the other axioms or from something else. We
have argued in this paper that, in the de Broglie-Bohm formulation of quantum
theory, the Born rule has a status similar to that of thermal equilibrium in
ordinary statistical mechanics, and should not be regarded as an axiom at all.
\\\\\\
{\Large Acknowledgement}\\
We wish to thank Rickard Jonsson for assistance with the figures.


\begin{thebibliography}{99}                                                         
\bibitem {deB}L. de Broglie, in: \textit{\'{E}lectrons et Photons: Rapports et
Discussions du Cinqui\`{e}me Conseil de Physique}, ed. J. Bordet
(Gauthier-Villars, Paris, 1928). [English translation: G. Bacciagaluppi and A.
Valentini, \textit{Electrons and Photons: The Proceedings of the Fifth Solvay
Congress} (Cambridge University Press, Cambridge, forthcoming).]

\bibitem {Bohm}D. Bohm, Phys. Rev. \textbf{85}, 166; 180 (1952).

\bibitem {Bell}J.S. Bell, \textit{Speakable and Unspeakable in Quantum
Mechanics} (Cambridge University Press, Cambridge, 1987).

\bibitem {AVth}A. Valentini, On the Pilot-Wave Theory of Classical, Quantum
and Subquantum Physics, PhD thesis, International School for Advanced Studies,
Trieste, Italy (1992).

\bibitem {Holl}P. Holland, \textit{The Quantum Theory of Motion: an Account of
the de Broglie-Bohm Causal Interpretation of Quantum Mechanics} (Cambridge
University Press, Cambridge, 1993).

\bibitem {BandH}D. Bohm and B.J. Hiley, \textit{The Undivided Universe: an
Ontological Interpretation of Quantum Theory} (Routledge, London, 1993).

\bibitem {Cush}J.T. Cushing, \textit{Quantum Mechanics: Historical Contingency
and the Copenhagen Hegemony} (University of Chicago Press, Chicago, 1994).

\bibitem {BM}\textit{Bohmian Mechanics and Quantum Theory: an Appraisal}, eds.
J.T. Cushing, A. Fine, and S. Goldstein (Kluwer, Dordrecht, 1996).

\bibitem {Duerr}D. D\"{u}rr, \textit{Bohmsche Mechanik als Grundlage der
Quantenmechanik} (Springer, 2001).

\bibitem {AVbook}A. Valentini, \textit{Pilot-Wave Theory of Physics and
Cosmology} (Cambridge University Press, Cambridge, forthcoming).

\bibitem {Pauli53}W. Pauli, in: \textit{Louis de Broglie: Physicien et
Penseur} (Albin Michel, Paris, 1953).

\bibitem {Keller}J.B. Keller, Phys. Rev. \textbf{89}, 1040 (1953).

\bibitem {Bohm53}D. Bohm, Phys. Rev. \textbf{89}, 458 (1953).

\bibitem {BV54}D. Bohm and J.P. Vigier, Phys. Rev. \textbf{96}, 208 (1954).

\bibitem {DGZ92}D. D\"{u}rr, S. Goldstein and N. Zangh\`{\i}, J. Stat. Phys.
\textbf{67}, 843 (1992); Phys. Lett. A \textbf{172}, 6 (1992).

\bibitem {PLA1}A. Valentini, Phys. Lett. A \textbf{156}, 5 (1991).

\bibitem {ValIsch}A. Valentini, in: \textit{Chance in Physics: Foundations and
Perspectives}, eds. J. Bricmont \textit{et al}. (Springer, Berlin, 2001) (quant-ph/0104067).

\bibitem {Berndl1}K. Berndl, D. D\"{u}rr, S. Goldstein, G. Peruzzi and N.
Zangh\`{\i}, Comm. Math. Phys. \textbf{173}, 647 (1995).

\bibitem {Berndl2}K. Berndl, in: \textit{Bohmian Mechanics and Quantum Theory:
an Appraisal}, eds. J.T. Cushing, A. Fine, and S. Goldstein (Kluwer,
Dordrecht, 1996).

\bibitem {Recipes}W.H. Press, S.A. Teukolsky, W.T. Vetterling and B.P.
Flannery, \textit{Numerical Recipes in FORTRAN: the Art of Scientific
Computing} (Cambridge University Press, Cambridge, 1992).

\bibitem {PV95}R.H. Parmenter and R.W. Valentine, Phys. Lett. A \textbf{201},
1 (1995).

\bibitem {PV96}R.H. Parmenter and R.W. Valentine, Phys. Lett. A \textbf{213},
319 (1996).

\bibitem {SF95a}U. Schwengelbeck and F.H.M. Faisal, Phys. Lett. A
\textbf{199}, 281 (1995).

\bibitem {Frisk}H. Frisk, Phys. Lett. A \textbf{227}, 139 (1997).

\bibitem {DM96}C. Dewdney and Z. Malik, Phys. Lett. A \textbf{220}, 183 (1996).

\bibitem {FS95b}F.H.M. Faisal and U. Schwengelbeck, Phys. Lett. A
\textbf{207}, 31 (1995).

\bibitem {Iac96}G. Iacomelli and M. Pettini, Phys. Lett. A \textbf{212}, 29 (1996).

\bibitem {Sen96}S. Sengupta and P.K. Chattaraj, Phys. Lett. A \textbf{215},
119 (1996).

\bibitem {PV97}R.H. Parmenter and R.W. Valentine, Phys. Lett. A \textbf{227},
5 (1997).

\bibitem {HW}H. Westman, unpublished.

\bibitem {Tol}R.C. Tolman, \textit{The Principles of Statistical Mechanics}
(Dover, New York, 1979).

\bibitem {Dav}P.C.W. Davies, \textit{The Physics of Time Asymmetry}
(University of California Press, Berkeley, 1974).

\bibitem {DSV}C. Dewdney, B. Schwentker and A. Valentini, in preparation.

\bibitem {Sklar}L. Sklar, \textit{Physics and Chance: Philosophical Issues in
the Foundations of Statistical Mechanics} (Cambridge University Press,
Cambridge, 1993).

\bibitem {Hall}\textit{Physical Origins of Time Asymmetry}, eds. J.J.
Halliwell, J. P\'{e}rez-Mercader, and W.H. Zurek (Cambridge University Press,
Cambridge, 1994).

\bibitem {Savitt}\textit{Time's Arrows Today: Recent Physical and
Philosophical Work on the Direction of Time}, ed. S.F. Savitt (Cambridge
University Press, Cambridge, 1995).

\bibitem {Pad}T. Padmanabhan, \textit{Structure Formation in the Universe}
(Cambridge University Press, Cambridge, 1993).

\bibitem {PLA2}A. Valentini, Phys. Lett. A \textbf{158}, 1 (1991).

\bibitem {AV96}A. Valentini, in: \textit{Bohmian Mechanics and Quantum Theory:
an Appraisal}, eds. J.T. Cushing, A. Fine, and S. Goldstein (Kluwer,
Dordrecht, 1996).

\bibitem {PLAc}A. Valentini, Phys. Lett. A \textbf{297}, 273 (2002).

\bibitem {Cracow}A. Valentini, in: \textit{Non-locality and Modality}, eds. T.
Placek and J. Butterfield (Kluwer, Dordrecht, 2002).

\bibitem {Sig}A. Valentini, quant-ph/0309107.

\bibitem {deWG}\textit{The Many-Worlds Interpretation of Quantum Mechanics},
eds. B.S. DeWitt and N. Graham (Princeton University Press, Princeton, 1973).

\bibitem {Deutsch}D. Deutsch, Proc. Roy. Soc. Lond. A \textbf{455}, 3129 (1999).

\bibitem {Wallace}D. Wallace, quant-ph/0211104; quant-ph/0303050; quant-ph/0312157.

\bibitem {Fuchs1}H. Barnum, C.M. Caves, J. Finkelstein, C.A. Fuchs and R.
Schack, Proc. Roy. Soc. Lond. A \textbf{456}, 1175 (2000).

\bibitem {Fuchs2}C.M. Caves, C.A. Fuchs and R. Schack, Phys. Rev. A
\textbf{65}, 022305 (2002).

\bibitem {G57}A.M. Gleason, J. Math. Mech. \textbf{6}, 885 (1957).

\bibitem {Bell66}J.S. Bell, Rev. Mod. Phys. \textbf{38}, 447 (1966).

\bibitem {SS}S. Saunders, Proc. Roy. Soc. Lond. A \textbf{460}, 1 (2004).

\bibitem {Zu}W.H. Zurek, Phys. Rev. Lett. \textbf{90}, 120404 (2003).

\bibitem {LH}L. Hardy, quant-ph/0101012; L. Hardy, in: \textit{Non-locality
and Modality}, eds. T. Placek and J. Butterfield (Kluwer, Dordrecht, 2002).

\bibitem {Bub}R. Clifton, J. Bub and H. Halvorson, Found. Phys. \textbf{33},
1561 (2003).
\end{thebibliography}
\end{document}